\documentclass{article}[12pt]
\usepackage{graphicx}
\begin{document}

\newcommand{\be}{\begin{equation}}
\newcommand{\ee}{\end{equation}}
\newcommand{\bea}{\begin{eqnarray}}
\newcommand{\eea}{\end{eqnarray}}
\newcommand{\bi}{\bibitem}
\newcommand{\la}{\langle}
\newcommand{\ra}{\rangle}
\renewcommand{\r}{({\bf r})}
\newcommand{\rp}{({\bf r'})}
\newcommand{\rpp}{({\bf r''})}
\newcommand{\rrp}{({\bf r},{\bf r}')}
\newcommand{\ua}{\uparrow}
\newcommand{\da}{\downarrow}
\newcommand{\s}{\sigma}

\title{\bf Spin-density-functional theory: some open problems and 
application to inhomogeneous Heisenberg models}
\author{K. Capelle and Valter L. L\'{\i}bero\\ \\
Departamento de F\'{\i}sica e Inform\'atica\\
Instituto de F\'{\i}sica de S\~ao Carlos\\
Universidade de S\~ao Paulo\\
Caixa Postal 369, 13560-970 S\~ao Carlos, SP\\
Brazil}
\date{\today}

\maketitle

\begin{abstract}
Spin-density-functional theory (SDFT) is the most widely implemented and
applied formulation of density-functional theory. However, it is still 
finding novel applications, and occasionally encounters unexpected problems. 
In this paper we first briefly describe a few of the latter, related to 
issues such as nonuniqueness, noncollinearity, and currents. In the main 
part we then turn to an example of the former, namely SDFT for the 
Heisenberg model. It is shown that time-honored concepts of
Coulomb DFT, such as the local-density approximation, can be applied
to this (and other) model Hamiltonians, too, once the concept of 'density'
has been suitably reinterpreted. Local-density-type approximations for the
inhomogeneous Heisenberg model are constructed. Numerical applications to
finite-size and impurity systems demonstrate that DFT is a computationally
efficient and reasonably accurate alternative to conventional methods of
statistical mechanics for the Heisenberg model.
\end{abstract}

\newpage

\section{\label{intro}Introduction}

Density-functional theory (DFT) is widely recognized as a versatile and 
powerful tool for practical calculations of the electronic structure of 
matter, on scales ranging from single atoms to extended solids. Its 
relatively low computational cost makes it attractive as a tool for 
answering material-specific questions in fields as diverse as quantum 
chemistry, materials science, and many-body physics \cite{kohnrmp}. In 
practice, DFT is most commonly used in its spin-dependent self-consistent-field
formulation as Kohn-Sham spin-density-functional theory (SDFT) 
\cite{ks,gunlund}. Alternative formulations that use other variables in 
addition to (or instead of) the spin densities, or that proceed without 
invocation of Kohn-Sham-type equations, are, however, occasionally useful. 
Some such alternative formulations are described in this contribution.
Section \ref{sdft} gives a brief overview of three recent developments of
this type in SDFT, relating to currents, noncollinear spin states and
nonuniqueness. In each case a surprising feature of the respective
extension of DFT will be highlighted. Section \ref{hemo} contains a 
more detailed description of a recent application of SDFT to model 
Hamiltonians of the Heisenberg type, where results can be obtained 
without solving Kohn-Sham equations.

\section{\label{sdft}Some recent developments in spin-density-func\-tional 
theory}

From a practical point of view, among the most important tasks remaining in 
the development of SDFT are to obtain ever better approximations for the 
exchange-correlation ($xc$) functional, and to learn how to calculate further
quantities of physical and chemical interest from the output quantities of 
SDFT, the spin-resolved charge densities. While total energies and related
quantities are readily accessible in terms of these densities, many other 
interesting quantities are not. 

\subsection{Electronic currents} 

Electronic currents are one example. Such currents are functionals of the 
charge density, but cannot be calculated as such because this functional is 
unknown. The current calculated from the Kohn-Sham (KS) orbitals of 
SDFT is that of the auxiliary noninteracting system, and not guaranteed by 
the basic theorems of SDFT to be the correct current. The full physical
(gauge invariant) current is ${\bf j}\r = {\bf j}_p\r+(e/mc)n\r {\bf A}\r$, 
where ${\bf A}\r$ is the vector potential and ${\bf j}_p\r$ is the so-called
paramagnetic current. The paramagnetic current calculated from the KS 
orbitals of SDFT is
\bea
{\bf j}^{KS}_p\r= \frac{\hbar}{2mi}\sum_k
\left[\varphi_k^*\r\nabla\varphi_k\r - \varphi_k\r\nabla\varphi_k^*\r\right]
\neq {\bf j}_p[n]\r.
\eea
Current-density functional theory (CDFT) \cite{vr1,vr2} provides a way to 
calculate the true equilibrium currents by obtaining the orbitals from a 
modified KS equation that features an $xc$ vector potential ${\bf A}_{xc}\r$, 
chosen such that in terms of these orbitals the above inequality becomes an 
equality. To achieve this, approximate $xc$ functionals of CDFT must depend 
on the current density, in addition to the charge and spin densities. This
dependence is a complicating factor both in the construction of approximate
functionals and in their implementation. In spite of these complications,
different ways of constructing CDFT functionals and solving the CDFT KS
equations have been explored for atoms, molecules and solids
\cite{vr1,vr2,ferconi,ebert,handy,jpe,rossler,pcdft}. As an alternative 
to solving the full CDFT equations, usable when current-dependent effects 
are expected to be small, it has been proposed to cast the CDFT KS equations 
in the form of SDFT KS equations plus a remainder depending on the $xc$ 
vector potential ${\bf A}_{xc}\r$, which is then treated perturbatively 
\cite{jpert}.

An unexpected feature of CDFT is that, although SDFT and CDFT are independent 
formulations of DFT for magnetic systems, there is a, somewhat obscure, link 
between CDFT and SDFT, which may be used to extract information on functionals 
of one from the other. In Ref.~\cite{roadprl} this link is obtained 
mathematically, but its physical origin is simple to understand:  
the existence of spin currents of the 
form ${\bf j}_s= \nabla \times {\bf m}$, where ${\bf m}$ is the 
spin magnetization, implies a connection between the formalisms of SDFT and 
CDFT, because being a manifestation of the spin degrees of freedom these
currents must be describable via SDFT, while as currents they can also be 
built into the formal framework of CDFT. Some consequences of the resulting 
connections and consistency relations between SDFT and CDFT are explored in 
\cite{roadprl}. For more details on CDFT we refer to 
Refs.~\cite{vr1,vr2,ferconi,ebert,handy,jpe,rossler,pcdft,jpert,roadprl}.

\subsection{Noncollinear magnetism}

The standard spin-up spin-down formulation of SDFT is incapable of describing 
spin configurations that are eigenstates of the square of the total spin 
operator $\hat{S}^2$, but not its $z-$component, $\hat{S}_z$.  Such 
noncollinear configurations can arise, e.g., in the presence of spin-orbit 
coupling, domain walls, external noncollinear fields, or from spontaneous 
symmetry breaking. Noncollinear ground states are found, e.g., in 
rare-earth based magnetic systems. In magnetic excitations, noncollinearity 
is ubiquitious. Consequently, noncollinear formulations both of static 
\cite{nordstroem,kubler,sandr} and time-dependent \cite{ziegler} DFT have 
been proposed, but applications are limited to the local-spin-density 
approximation (LSDA). In trying to describe noncollinear spin states with
generalized-gradient approximations (GGAs) or other beyond-LSDA functionals 
an interesting problem arises.

Due to its origin in the electron liquid, the exchange-correlation energy 
in the LSDA depends only on the modulus of the spin magnetization ${\bf m}\r$. 
The corresponding conjugate field, the $xc$ magnetic field 
\be 
{\bf B}_{xc}^{LSDA}\r=
-\frac{\delta E^{LSDA}_{xc}[n,|{\bf m}|]}{\delta {\bf m}\r},
\ee
is then always parallel to ${\bf m}$.
Beyond the LSDA, however, the $xc$ magnetic field ceases to be parallel to the 
spin magnetization. Mathematically this occurs because the existence of 
spatial derivatives of any component of ${\bf m}$ in the functional leads 
to additional terms in the derivative that are not parallel to ${\bf m}$ 
\cite{spindynprl,gilbert}. Intuitively this can be understood because the
existence of spatial gradients of the spin density introduces another
preferred direction into the problem. In general, the $xc$ magnetic field 
can thus be decomposed as
\be
{\bf B}_{xc}\r= {\bf B}_{xc}^{||}\r + {\bf B}_{xc}^{\perp}\r,
\ee
where the second contribution is locally perpendicular to the spin 
magnetization ${\bf m}\r$. This perpendicular $xc$ magnetic field exerts
an $xc$ torque $\propto {\bf m}\r \times {\bf B}_{xc}\r$ on the spin 
distribution. In the time-dependent case some consequences of this torque 
were described in \cite{spindynprl}, but it is present even in the ground 
state as soon as noncollinearity appears and is treated beyond the LSDA. 
This subtle effect is little explored in present-day SDFT.

Standard implementations of noncollinear SDFT assume parallelity between 
${\bf B}_{xc}$ and ${\bf m}\r$, and can thus not be used consistently with 
GGA-type functionals. What the best way is to treat noncollinearity beyond 
LSDA is an open question at present. As an alternative to formulations in 
terms of the spin magnetization, it has been proposed, e.g., to employ the 
spin-offdiagonal element of the density matrix (the so called 'staggered 
density' \cite{stagg1}), but this requires the construction of functionals 
of this new variable, and the solution of more complicated KS equations 
\cite{stagg2}.

\subsection{Nonuniqueness}

At first sight more remote from numerical applications are fundamental
issues of SDFT, such as the recently discussed question of
nonuniqueness of the SDFT potentials. The question to what extent
the effective potentials of SDFT (and other generalized DFTs) are uniquely
determined by the densities  arises because contrary to what is
sometimes claimed in the literature, the standard proofs of the Hohenberg-Kohn
theorem (by contradiction \cite{hk} and by constrained search \cite{levy}) 
guarantee only the existence of a one-on-one mapping between densities and 
wave functions, not densities and potentials. 
The extension of the mapping to potentials requires, as an additional step, 
inversion of Schr\"odinger's equation to express the potential in terms of 
the universal operators $\hat{T}$ and $\hat{U}$ and the wave function
\cite{levy,dftbook}. For charge-only DFT one has simply
\be
\hat{V}=E_0 - \frac{(\hat{T}+\hat{U})\Psi_0}{\Psi_0},
\ee
which shows that the ground-state wave function $\Psi_0$ (which is known to be
a functional of the density) determines the potential $V$ uniquely up to an
additive constant (here the ground-state energy, $E_0$). Since one always 
has an additive constant free in the definition of a potential, this shows
that the physically relevant part of $\hat{V}$, in particular its variation 
in space, is fully determined by $\Psi_0$ and hence by $n\r$.
Already in the early literature on noncollinear SDFT it was pointed out that 
this inversion is problematic if one works with 2-component spinors and has 
four potential components $(v,B_x,B_y,B_z)$ to determine. Von Barth and Hedin
gave an explicit example of the resulting nonuniqueness of the SDFT effective
potentials, albeit one that only works for a single electron 
\cite{vbarthhedin}.

Recently, examples have been given that show that nonuniqueness is not a
pathological feature of single-electron systems, or limited to SDFT, but a
general phenomenon, expected to occur whenever one works with more than one
density variable \cite{ep,nonunprl}. 
In Ref.~\cite{nonunprl} it was shown how a large class of nonunique potentials 
is related to the existence of certain types of constants of motion 
(systematic nonuniqueness) while others arise from special features
of the ground state (accidental nonuniqueness). A few simple examples from
collinear SDFT were given, illustrating the main features of both types of 
nonuniqueness. The simplicity of these examples notwithstanding, their 
existence shows that there cannot be a unique mapping between densities
and potentials in generalized DFTs, in addition to the one between
densities and wave functions. More complex examples
of nonuniqueness were found to be possible in noncollinear situations
\cite{ep,nikitas}, on lattices \cite{carsten} or in CDFT \cite{nonunprb}.
The implications of nonuniqueness for practical SDFT calculations are still
under study \cite{ep,nikitas,carsten,argaman,kohn}.
One conclusion that has emerged from this body of work is that the additional 
mapping between density and potential must be regarded as a special feature 
of spin unpolarized (charge-only) DFT. Surprisingly, the HK theorem of 
unpolarized (charge-only) DFT is thus considerably stronger than its 
counterparts in, e.g., SDFT and CDFT. 

\section{\label{hemo}Spin density-functional theory for the Heisenberg model}

Open problems, such as the best way to deal with currents, noncollinearity
and nonuniqueness, provide fascinating challenges, but they have never stoped
DFT from advancing on the practical side. 
In a typical application of DFT one deals with the {\it ab initio} 
many-electron Hamiltonian comprising kinetic, potential, and (Coulomb) 
interaction terms. By means of the Hohenberg-Kohn (HK) theorem and the 
Kohn-Sham construction this many-body problem is mapped on an auxiliary 
noninteracting one. The search for good 
approximations for the exchange-correlation potential appearing in the 
auxiliary KS Hamiltonian is one of the most active fields of DFT. However, the
HK and KS techniques are not restricted to the {\it ab initio} Hamiltonian. 
We can simply regard them as efficient tools for mapping {\it some} many-body
problem on a simple one-body problem. In many areas of science, most notably
perhaps in statistical physics and many-body physics, but occasionally also in 
quantum chemistry, the fundamental Hamiltonian is replaced by a simpler model
Hamiltonian before embarking on any numerical calculation. The list of such 
simplified Hamiltonians is long, and comprises important models such as those 
of Ising, Heisenberg and Hubbard. This section is devoted to a description of 
a recent reformulation and application of SDFT to model Hamiltonians of the 
Heisenberg type \cite{nos1,nos2}. 

\subsection{The Heisenberg model}

Introduced in 1928 by Dirac and Heisenberg \cite{dirac,heis} and discussed 
in some detail as a model for ferromagnetic solids in 1932 by van Vleck 
\cite{van}, the Heisenberg model is the paradigmatic model for the magnetism 
of local magnetic moments ${\bf S}_i$. In quantum chemistry it has, e.g., 
been used in the description of the magnetic properties and spin states of 
conjugated hydrocarbons \cite{malrieu1,malrieu2} and a large variety of 
metallic and organometallic compounds \cite{tchougreeff,kahn,kityk,luscombe}.

In its simplest form the Heisenberg model reads
\be
\hat{H} = J \sum_{\la ij\ra} \hat{\bf S}_i \cdot \hat{\bf S}_j \;,
\label{heisen}
\ee
where the $\hat{\bf S}_i$ are angular-momentum vector operators
satisfying $\hat{\bf S}^2|Sm\ra = \hbar^2 S(S+1) |Sm\ra$. 
In accordance with common terminology we refer to the expectation values 
of $\hat{\bf S}_i$, the magnetic moments ${\bf S}_i$, as spins, although they 
may also have an orbital contribution. The sums run over nearest neighbours 
among the $N$ sites of a lattice of, in principle, arbitrary geometry, and
$J$ parametrizes the nearest-neighbor
interactions. We see qualitatively that a parallel (ferromagnetic) or 
antiparallel (antiferromagnetic) alignment of ${\bf S}_i$ and ${\bf S}_j$ is 
favored according to whether $J$ is negative or positive. 

In the presence of an external magnetic field ${\bf B}_i$ the model takes the 
form
\be
\hat{H({\bf B})} = J \sum_{\la ij\ra} \hat{\bf S}_i \cdot \hat{\bf S}_j 
+ \sum_i \hat{\bf S}_i \cdot {\bf B}_i\;,
\label{heisenB}
\ee
If ${\bf B}_i$ and ${\bf S}_i=\la \hat{\bf S}_i\ra$ are the same everywhere,
the model is homogeneous, i.e., all its sites are equivalent.
In all other cases it is inhomogeneous. Note that even for ${\bf B}_i=0$
(no external field) inhomogeneous states are possible. Such solutions may 
arise because the self-consistent many-body ground state need not preserve all
symmetries of the Hamiltonian, or because the Hamiltonian itself comprises
inequivalent lattice sites. The latter is the case, e.g., in systems with
impurities, or, more generally, with of different types of magnetic
atoms.

In spite of its formal simplicity, exact analytical results for the Heisenberg 
model are known only for homogeneous linear chains with $S=1/2$, by means of
the Bethe ansatz \cite{bethe,hulthen}. For higher spin, higher dimension, or
inhomogeneous situations, exact solutions are known only numerically, for 
small systems (less than $\approx 40$ sites), where the model can be 
diagonalized exactly by Lanczos techniques. In the presence of 
inhomogeneities, such as one substitutional spin of value $S' \neq S$, no 
analytical solution is known. 

The simplest approximation to the Hamiltonian (\ref{heisen}) is the
mean-field (Hartree-like) approximation, in which the vector operators 
$\hat{\bf S}$ are substituted by classical vector spins ${\bf S}$. 
The mean-field ground-state energy per spin, in units of $J$, is thus 
easily obtained for a homogeneous infinite system of spins $S$ in $d$ 
dimensions (with periodic boundary conditions):
\begin{equation}
e_0^{MF}(S,d) = \frac{E_0^{MF}(S,d)}{NJ}= -d\;S^2 \;,
\end{equation}
where $d$ is the dimensionality of the linear ($d=1$), square ($d=2$)
or cubic ($d=3$) lattice. Other lattice geometries can be dealt with in 
the same way.
An improvement on the mean-field estimate for $e_0(S,d)$ is obtained from 
spin-wave theory \cite{spinwave}, according to which
\begin{eqnarray}
e_0^{SW}(S,d) = \frac{E_0^{SW}(S,d)}{NJ} = 
-d\;S^2 + d^{-1/5} \left(\frac{2}{\pi}-1\right)S \;.
\end{eqnarray}
Here we used the scaling hypothesis conjectured in Ref. \cite{nos1}, which 
accounts for the numerical spin-wave results to three decimal places, to write 
the results in different dimensionalities in closed form, as a function of $d$.
Beyond spin-wave theory, sophisticated numerical many-body techniques, such
as the density-matrix renormalization group (DMRG) \cite{lou} or Quantum Monte 
Carlo simulations \cite{qmc}, have been employed to further improve the 
ground-state energy.

The resulting expressions for $E_0$, however, suffer from a major limitation,
namely, they are applicable only to homogeneous systems, in which all lattice
sites are equivalent. This limitation is restrictive, because in real 
problems it is very common to find
inhomogeneities, induced for instance by external non-uniform magnetic 
fields, anisotropic crystal-fields, finite-size effects, boundary 
conditions, local impurities, lattice defects, etc. These rather common 
situations introduce very significant difficulties in traditional 
approaches like Bethe ansatz, DMRG, spin-wave theory or even
Monte Carlo, by breaking translational symmetry. In order to make progress
in dealing with {\em inhomogeneous} Heisenberg models we need a many-body 
technique capable of handling spatial inhomogeneity in large systems. 
Density-functional theory is our first choice in this regard.

\subsection{Spin-distribution functionals and local-spin approximation}
\label{inhom}

Density-functional theory \cite{kohnrmp,dftbook} offers, with the local-density
approximation (LDA), a simple prescription for obtaining ground-state 
properties of an inhomogeneous system, based on the knowledge of a related 
homogeneous system. While in {\it ab initio} applications of DFT this 
homogeneous system is the electron liquid, in our context it is the 
homogeneous Heisenberg model, in which all sites are equivalent.

Ref. \cite{nos1} shows that the basic ingredients of DFT (in particular the 
Hohenberg-Kohn theorem \cite{hk}) stilll hold for Hamiltonians of the form
(\ref{heisenB}), provided one uses instead of the ground-state density $n\r$
the ground-state expectation value of the spin vectors, 
${\bf S}_i=\langle \Psi|{\bf S}_i|\Psi \rangle$.
The Heisenberg-model Hohenberg-Kohn theorem then states that the ground-state 
expectation value of any observable $\hat{O}$ is a functional of the
distribution ${\bf S}_i$, i.e., $O[{\bf S}_i]= \la \Psi[{\bf S}_i] | \hat{O} | \Psi[{\bf S}_i] \ra$. In this Heisenberg-model DFT the above mean-field
expressions take the place of the Hartree term in Coulomb DFT, and
the correlation energy $e_c(S)$ is defined as the difference
$e_0(S)-e_0^{MF}(S)$, where $e_0(S)$ is the full ground-state energy.
From the above, the spin-wave approximation for $e_c(S)$ is thus
\be
e_c^{SW} = e_0^{SW} - e_0^{MF} =  d^ {-1/5}\left(\frac{2}{\pi}-1\right)S \;.
\ee

The Heisenberg-model Hohenberg-Kohn theorem, proved in \cite{nos1}, 
guarantees that in inhomogeneous systems the total energy is a functional
of the spatial distribution of the classical vectors ${\bf S}_i$. Hence
\be
E_0=E_0[{\bf S}_i] = E^{MF}_0[{\bf S}_i] + E_c[{\bf S}_i] \;.
\ee
Since the mean-field term is already a functional of ${\bf S}_i$,
we just need an approximation for the functional $E_c[{\bf S}_i]$. This
is obtained using a {\em local-spin approximation} (LSA), patterned after 
the conventional LDA of Coulomb DFT. The LSA consists in replacing
locally the variable $S$ by $S_i$ in $e_c(S)$:
\begin{equation}
E_c[{\bf S}_i] \approx
E_c[{\bf S}_i]^{LSA} = \sum_i^N e_c(S)|_{S\to |{\bf S}_i|} \;.
\end{equation}
Using the above spin-wave approximation for $e_c(S)$ we obtain, e.g.,
\begin{equation}
E_c^{LSA-SW}[{\bf S}_i,d,J]= 
d^{-1/5} J\left(\frac{2}{\pi}-1\right)\sum_{i=1}^N |{\bf S}_i| \;,
\end{equation}
and thus the ground-state energy functional in this approximation is
\begin{equation}
E_0^{LSA-SW}[{\bf S}_i,d,J] =
J \sum_{\la ij\ra} {\bf S}_i \cdot {\bf S}_j +
d^{-1/5} J \left(\frac{2}{\pi}-1\right) \sum_i |{\bf S}_i| \;.
\label{swlsa}
\end{equation}
More sophisticated expressions, based on fully numerical evaluation of
$e_0(S)$ are also available \cite{nos1}, but in first applications 
the simple approximation (\ref{swlsa}) turned out to be already reasonably
accurate \cite{nos2}.

As always in dealing with LDA-type functionals one must distinguish two 
sources of error. One is the LDA itself, the other the quality of the
description of the underlying homogeneous reference system. In Coulomb
DFT several alternative parametrizations of the LDA are available 
\cite{vbarthhedin,glvwnpzpw}, which differ considerably in form, and
slightly in results. In the present context, too, one can construct
various LSA functionals, depending on the level of sophistication employed
in solving the uniform Heisenberg model. For example,
for an infinite linear chain of spin $S=1/2$ we have from the above 
spin-wave based LSA $e_0/(JN) = -0.25-0.181690 = -0.431690$, which is 
$2.6\%$ from the exact Bethe ansatz value $-0.443147$. 
Interestingly, we find this same margin of deviation also when the LSA 
is used for inhomogeneous situations (see below). This seems to indicate
that the LSA concept as such is quite reliable for Heisenberg models.

From the above, it should have become clear that the formalism of DFT for
the Heisenberg model is built in complete parallelity to that of
Coulomb DFT (or that for the Hubbard model \cite{balda}).
One interesting difference is that within spin-wave theory the 
dimension dependence of the functional is approximately known. This is very
different from the {\it ab initio} case, in which nothing is known about
this dependence. In general, however, one may expect that LSA for the
Heisenberg model shares the virtues and defects of its close cousin, the
LDA for the Coulomb problem: efficient access to large and inhomogeneous
systems on one side, offset by moderate accuracy on the other. 
In the next subsection we demonstrate these aspects by analysing two
interesting inhomogeneous models within LSA. (Other applications of the 
above functionals can be found in Ref. \cite{nos2}.)

\subsection{Applications to inhomogeneous Heisenberg models: boundaries
and impurities}

The usual DFT prescription for obtaining ground-state energies is to solve
Kohn-Sham equations. These equations, however, were originally \cite{ks}
introduced to deal with the kinetic-energy term in the Coulomb Hamiltonian.
The Heisenberg model does not have such a term, and direct minimization 
of the energy functional is much simpler than indirect minimization via
self-consistency (Kohn-Sham) equations. 
In practice we just need to minimize the total energy of the system,
written as a functional of the spin distribution, with respect to 
${\bf S}_i$. This already represents a considerable improvement over 
the mean-field approximation with no extra computational cost; minimization of
$E_0^{LSA-SW}[{\bf S}_i]$ is no more complicated than that of 
$E_0^{MF}[{\bf S}_i]$.

\begin{figure}
\includegraphics[height=200mm,width=140mm,angle=0]{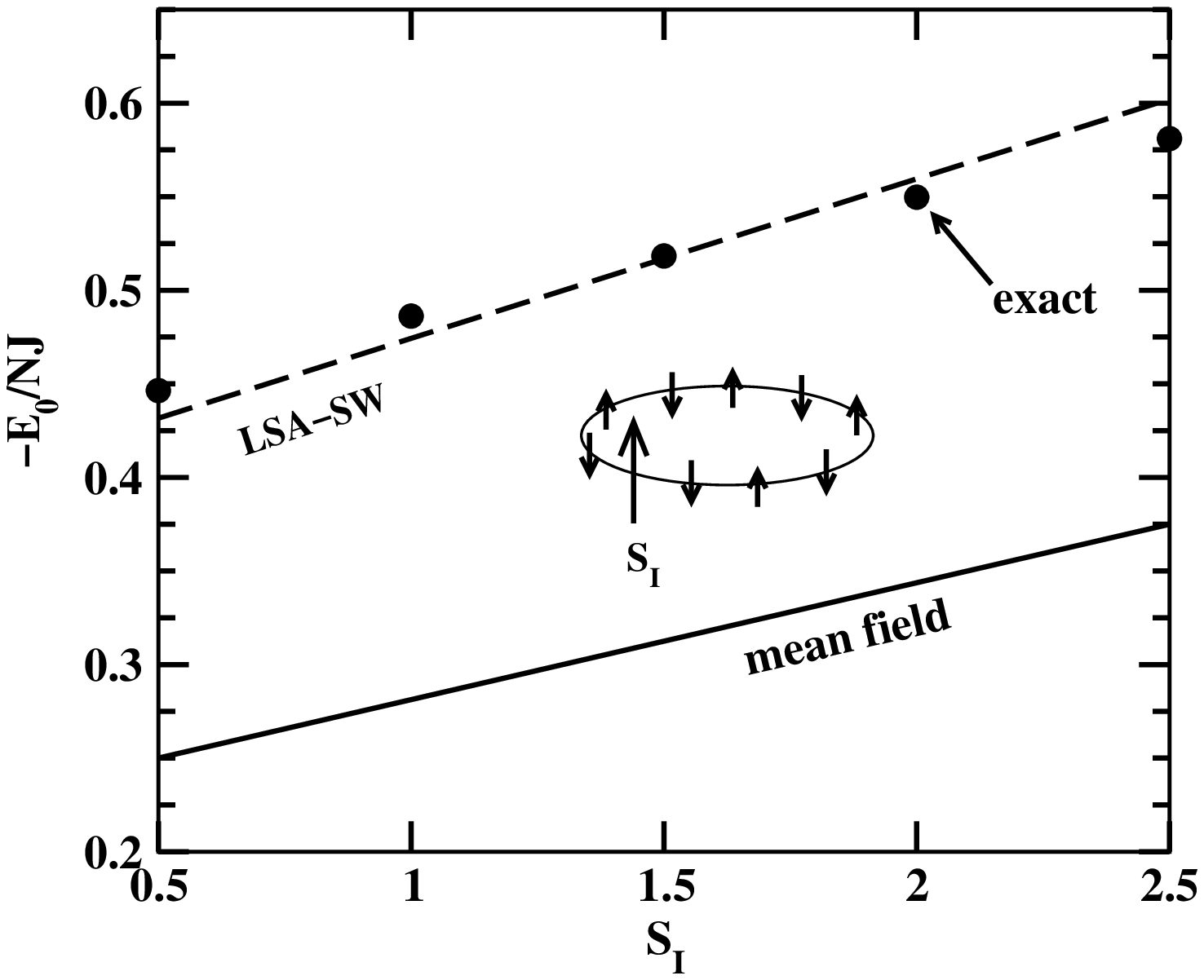}
\caption{Per-site ground-state energy of a ring of $N=16$ spin $S=1/2$ sites 
as a function of the impurity spin $S_I$. The LSA results, based on spin-wave
theory, (dashed line) deviate from the exact ones (circles) by about $3\%$.
The mean field approximation (continuous line) deviates by up to $50 \%$. The
inset illustrates the geometry under study, for the case of $N=10$.}
\label{fig1}
\end{figure}

Lets first consider a finite ring of $N=16$ spins $S= 1/2$, with periodic 
boundary conditions. The mean field approximation yields $e_0=-0.25$, while 
the $LSA-SW$ approximation gives $e_0=-0.43169$, about $3\%$ from the exact 
value $-0.44639$ obtained by direct diagonalization. Let us now replace any 
spin of the ring by an impurity spin $S_I$. This slight modification is enought
to challenge any analytical approach, but since the ring is not too big, it 
is still possible 
to numerically diagonalize $\hat{H}$ in order to get the exact ground 
state. Figure \ref{fig1} shows how the ground-state energy depends on $S_I$, 
and compares the values obtained by mean-field and LSA-SW calculations with
the numerically exact ones. Up to $S_I=2.5$ the deviation from the exact 
values are of order $3 \%$. Above this value the deviation increases, showing
that the LSA approximation is not good for large impurity spin (corresponding 
to a very rapid variation of local magnetic properties). For impurity spins 
up to $S_I=5/2$, on the other hand, LSA-SW is seen to provide a substantial
improvement on the mean-field approximation. Unlike the exact calculations,
LSA-SW can easily be extended to systems with more impurities and more
sites. It may thus provide a convenient way of estimating (and improving
on) the error of the mean-field approximation in more complicated systems.

Next we consider $4\times 4$ square lattices with an impurity of 
spin $S_I$ at the corner, at one side, or in the interior of the lattice. 
Here we use open boundary condition and thus even without impurity the systems
are inhomogeneous. Results are collected in Tab.~\ref{table1}. The first 
column is for $S_I=1/2$, i.e., for 
the uniform system, and the value listed differs $6.6\%$ from the exact value 
$e_0=-0.574325$ obtained from numerical diagonalization; we believe the values
for other impurity spins are within this accuracy, although no exact values 
are available, as far as we know. Besides the obvious energetic tendency of 
higher impurity-spin to prefer the interior of the lattice to its surface, 
which is already visible in the mean-field results, these data also incorporate
less trivial correlation effects: In mean-field theory, for example, a
spin 3/2 at the corner of the square is degenerate with a spin 1 in the 
interior of the lattice ($e^{MF}_0=-0.4375$ in both cases). This degeneracy 
is broken by correlation effects. Similarly, mean-field theory predicts 
degeneracy between a spin 5/2 at one side and a spin 2 in the interior 
($e^{MF}_0=-0.5625$ for both cases). Again this degeneracy is removed by 
including correlations. 

\section{Summary}

Kohn-Sham spin-density functional theory is doubtlessly the most widely used 
formulation of density-functional theory. However, in spite of its great 
popularity and usefulness some interesting fundamental questions still remain 
open. The existence of spin currents, providing a link between SDFT and CDFT;
the existence of an exchange-correlation torque appearing in noncollinear
spin configurations; and the nonuniqueness of the potentials of SDFT and other
generalized DFTs are examples of the kind of surprise the SDFT formalism
still holds.

Such open questions notwithstanding, the utility of (S)DFT is extending
even into areas remote from {\it ab initio} calculations. Recognition that
the basic tools and concepts of DFT --- such as the Hohenberg-Kohn theorem,
and the local-density approximation --- are not restricted to the original
Coulomb problem allows one to apply these tools and concepts also to many
other interesting inhomogeneous model Hamiltonians. In particular,
density-functional theory together with results from mean-field and spin-wave
theory provide a simple manner to obtain estimates of the ground-state 
energy for spin distributions described by the inhomogeneous Heisenberg 
model. This is here illustrated by results for a finite ring and a 
$4 \times 4 $ lattice, both with substitutional impurity spins $S_I=1,3/2,2$ 
or 5/2. While ring calculations may find applications in modelling the spin 
states of hydrocarbons, the square-lattice data illustrate a way in which 
LSA can be useful in nanomagnetism: To predict the structure of self-assembled
magnetic nanostructures it is clearly important to know which lattice sites 
are degenerate in mean-field theory, and which of these degeneracies are 
removed by correlations.

{\it
This work was supported by FAPESP and CNPq. We thank P.~E.~G.~Assis, and
F.~C.~Alcaraz for useful discussions and collaboration on the Heisenberg
model. K.C. thanks L.~N.~Oliveira, E.~K.~U. Gross and G.~Vignale for useful 
discussions and collaborations on various aspects of spin-density-functional 
theory.}

\newpage

\begin{table}
\caption{\label{table1} LSA-SW ground-state energy $-E_0/NJ$ of a 
$4 \times 4$ square lattice of $S=1/2$ sites with a substitutional impurity 
of spin $S_I$ at the corner (four equivalent positions), on a side (eight 
equivalent positions) or in the interior of the lattice (four equivalent 
positions).\protect\\ }
\begin{tabular}{r|c|c|c|c|c}
         & $1/2$ & 1 & 3/2 & 2 & 5/2  \\
         &       &   &     &   &      \\
\hline
corner   & 0.612 & 0.653 & 0.694 & 0.735 & 0.776 \\
side     & 0.612 & 0.669 & 0.726 & 0.782 & 0.839 \\
interior & 0.612 & 0.684 & 0.757 & 0.829 & 0.902
\end{tabular}
\end{table}

\newpage

\end{document}